\documentclass{iau}

\usepackage{amsmath}
\usepackage{graphicx}
\usepackage{multirow}

\newcommand{\HI}{H\,{\sc i}}

\newcommand{\kms}{km\,s$^{-1}$}

\newcommand{\Htwo}{H$_{2}$}

\newcommand{\co}{CO}

\newcommand{\hix}{H\,{\sc ix}}

\begin{document}

\lefttitle{D{\'e}nes, H and Capa, V. A.}
\righttitle{The mystery of extremely HI rich galaxies}

\jnlPage{1}{7}
\jnlDoiYr{2021}
\doival{10.1017/xxxxx}

\aopheadtitle{Proceedings IAU Symposium 392}
\editors{D.J. Pisano, M. Mogotsi, Julia Healy, \& S. Blyth}

\title{The mystery of extremely HI rich galaxies}

\author{D{\'e}nes, H. and Capa, V. A.}
\affiliation{School of Physical Sciences and Nanotechnology, Yachay Tech University, Hacienda San Jos{\'e} S/N, 100119, Urcuqu{\'i}, Ecuador}

\begin{abstract}
The properties of galaxies follow scaling relations related to the physics that govern galaxy evolution. Based on these, we can identify galaxies undergoing specific evolutionary processes such as \HI-excess galaxies, which have relatively high \HI\ mass compared to their stellar mass. The possible reasons for this could be either recent gas accretion or an inefficient conversion of the \HI\ to molecular gas. Since recent gas accretion is difficult to prove conclusively, we investigated gas conversion by analysing the molecular gas content of five extremely \HI\ rich galaxies from the \hix\ galaxy sample. For this, we obtained \co\ observations of the sample galaxies with the Atacama Large Millimeter Array (ALMA). While we find that our sample galaxies have relatively regularly rotating molecular gas disks, their molecular gas fraction is significantly lower than what is expected from scaling relations and the \co\ gas has relatively high velocity dispersion. 
\end{abstract}


\maketitle

\section{Introduction}
 
Typical late-type, star forming galaxies tend to have large neutral hydrogen (\HI) reservoirs distributed in a disc. The \HI\ content of late-type galaxies correlates with their optical properties, such as the luminosity, diameter and specific star formation rate (e.g. \citealt{Catinella2012, Denes2014}). However, due to various physical processes, such as gas stripping and gas accretion, there is a substantial scatter on these scaling relations and several galaxies are significantly removed from the relations. While the \HI-deficient population has been extensively studied, we are only starting to understand \HI-rich galaxies.

Motivated by this, the \hix\ survey \citep{Lutz2017,Lutz2018} mapped the \HI\ content of 12 nearby \HI-eXtreme galaxies with the Australia Telescope Compact Array (ATCA) to investigate the reason for their \HI-richness. These galaxies were selected to contain at least 2.5\,times more \HI\ than expected based on the $R$-band \HI\ mass scaling relation \citep{Denes2014}. The \hix\ survey galaxies look like average star forming spirals, based on their stellar mass, but they reside in halos with higher than average angular momentum (spin) \citep{Lutz2018}. Thus, they are able to build up and maintain large \HI\ discs, with \HI\ masses up to $\sim 10^{10.8}$M\,$_\odot$ and \HI\ disc sizes 2.3 to 5 times the 25\,mag\,arcsec$^{-2}$ isophotal radius \citep{Lutz2018}. Due to the large amount of \HI\ located outside the stellar disc, \hix\ galaxies have a lower global, \HI-based star formation efficiency than average late-type galaxies of similar mass. 

The higher average angular momentum of these galaxies could also affect their molecular gas. For typical nearby galaxies the xGASS and xCOLD GASS \citep{Saintonge2017,Catinella2018} surveys provide insights into the relation between the global \HI\ and \Htwo\ properties. These studies found that the \Htwo\ to \HI\ ratio increases both with stellar mass and with morphological type, when moving from disc to bulge dominated systems \citep{Catinella2018}. Studies of spatially resolved nearby galaxies also found that the the \Htwo\ to \HI\ fraction correlates with the stellar surface brightness and anti correlates with the star formation efficiency (e.g. \citealt{Bigiel2016}). This implies that the \Htwo\ fraction is a limiting factor in the star formation efficiency. However, a study of three nearby \HI-rich galaxies by \cite{Hallenbeck2016} found that their sample has typical \Htwo\ disks and star formation rates, which suggests that these galaxies are inefficient in converting their \HI\ into \Htwo\ rather than being inefficient at forming stars. 

To investigate which processes are relevant for the \hix\ galaxies, we mapped the molecular gas content of 5 \hix\ galaxies by observing the carbon monoxide CO(1-0) rotational transition. Here we present the molecular gas properties of these galaxies and compare them to the \HI\ and stellar mass scaling relations. 

\section{Data}

We obtained ALMA \co\ line data for 5 \hix\ galaxies: ESO208-G026, ESO378-G003, ESO381-G005, ESO417-G018, ESO055-G013, under the ALMA project code: 2019.2.00169.S. All five galaxies have been observed with the 7m array, and the three more extended galaxies (ESO378-G003, ESO381-G005 and ESO417-G018) have been observed with the total power array as well. ALMA observations were taken in band 3 assuming the same systemic velocity for the \co\ as for the \HI. We used a frequency resolution of $\sim$ 1.1290 MHz which corresponds to a velocity resolution of 2.6 \kms. 

To improve the signal-to-noise of our data cubes, we also created binned data cubes by binning every 5 channels, which resulted in a velocity resolution of 14 \kms. We used {\sc SoFiA} (Source Finding Application, \citealt{Serra2015, Westmeier2021}) to identify the signal in both the original and binned data cubes to create masked cubes with only the reliably detected \co\ signal. In addition to the masked cube, we also obtained the integrated intensity, velocity and velocity dispersion maps.

\begin{figure}[t]
  \centerline{\vbox to 6pc{\hbox to 10pc{}}}
  \includegraphics[width = 7cm]{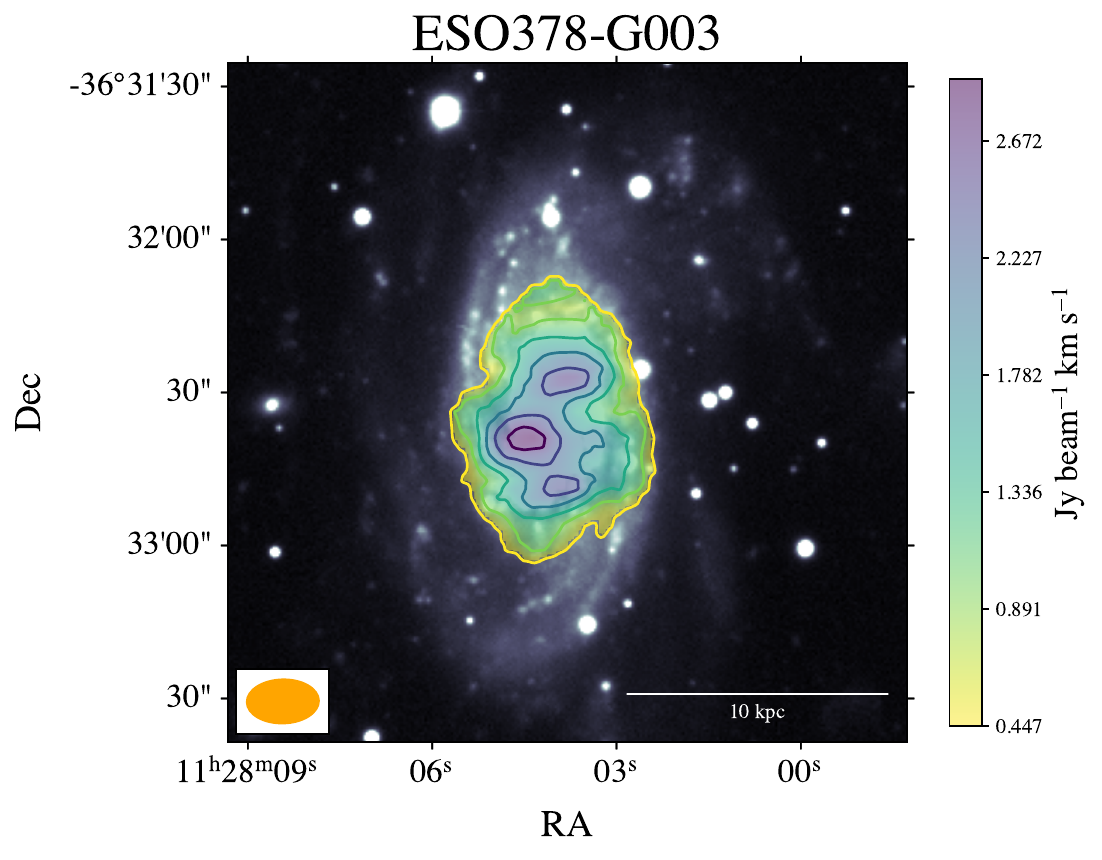}
  \includegraphics[width = 7cm]{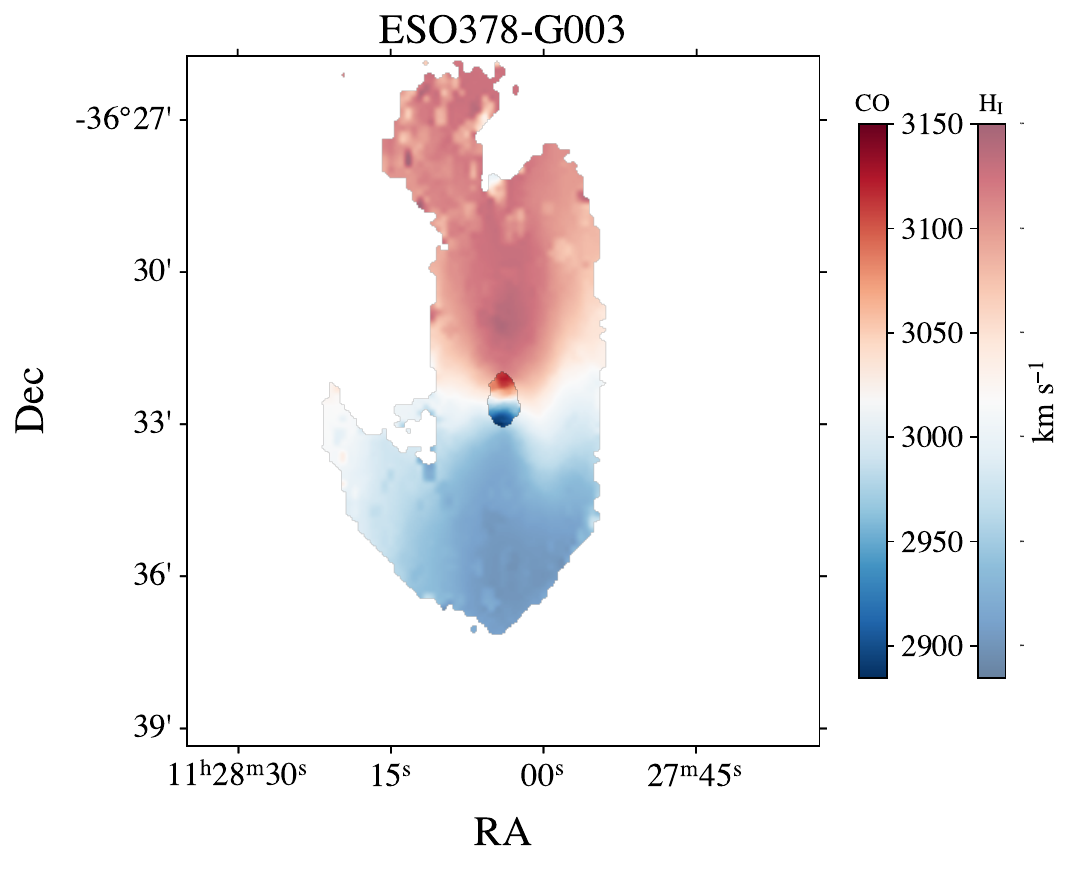}
  \caption{Left: Total CO intensity map for ESO 378-G003 overlayed on the Legacy Survey DR10 g-band optical image of the galaxy. The CO beam is indicated in the lower left corner. Right: \HI\ (lighter colour) and CO (darker colour) intensity weighted velocity field for the same galaxy.}
  \label{fig:Maps}
\end{figure}

\section{Molecular gas properties}

Figure~\ref{fig:Maps} shows an example of our \co\ data set with the total \co\ intensity map for ESO 378-G003 overlayed on the Legacy Survey DR10 g-band optical image\footnote{\url{https://www.legacysurvey.org/}} and an overlay of the \HI\ and \co\ intensity weighted velocity fields. Similar to typical late-type galaxies, the molecular gas disk is covering the inner parts of the stellar disk and is substantially smaller compared to the \HI\ disk. The velocity field of the molecular gas agrees well with the velocity field of the \HI.

Interestingly, we found relatively high \co\ velocity dispersion in all of the sample galaxies, with distributions peaking at 10-15 \kms\ and maximum values around 55\kms. We observed the lowest \co\ velocity dispersion for ESO378-G003, which seems to have the most regular \Htwo\ content compared to its stellar mass. Compared to the measurements of nearby galaxies presented in \cite{Mogotsi2016}, we found that the \hix\ sample shows higher velocity dispersion, which could indicate high turbulence in the \hix\ galaxies. 

\subsection{H$_2$ mass and scaling relations}

From the flux density maps of \co\ gas, we calculated the \Htwo\ mass of each galaxy, using the equation from \citealt{Accurso2017},

\begin{equation}
    M_{H_2} = \alpha_{CO} L'_{CO(1-0)}
    \label{eq:Accurso}
\end{equation}
\begin{equation}
    L'_{CO(1-0)} = 3.25 \times 10^7 I_{CO}\Delta v \nu_{obs}^{-2} D_{L}^2 (1+z)^{-3}
    \label{eq:Accurso}
\end{equation}

where $\alpha_{CO}$ is the \co\ to \Htwo\ conversion factor, $L'_{CO(1-0)}$ is the integrated \co\ line luminosity in K km s$^{-1}$ pc$^2$, $I_{CO(1-0)}\Delta v$ is the the observed velocity integrated flux density in Jy km s$^{-1}$, $D_{L}$ is the luminosity distance in Mpc, $\nu_{obs}$ is the observed frequency in GHz and $z$ is the redshift. We used $\alpha_\text{\co} = 4.36\,M_{\odot} \rm{(K\,km\,s^{-1})}^{-1}$, to be able to compare our datasets to the xCOLD GASS galaxy sample\footnote{\url{http://www.star.ucl.ac.uk/xCOLDGASS/index.html}} (\citealt{Saintonge2017}, \citealt{Accurso2017}). 

\begin{figure}[t]
  \centerline{\vbox to 6pc{\hbox to 10pc{}}}
  \includegraphics[width = 6.5cm]{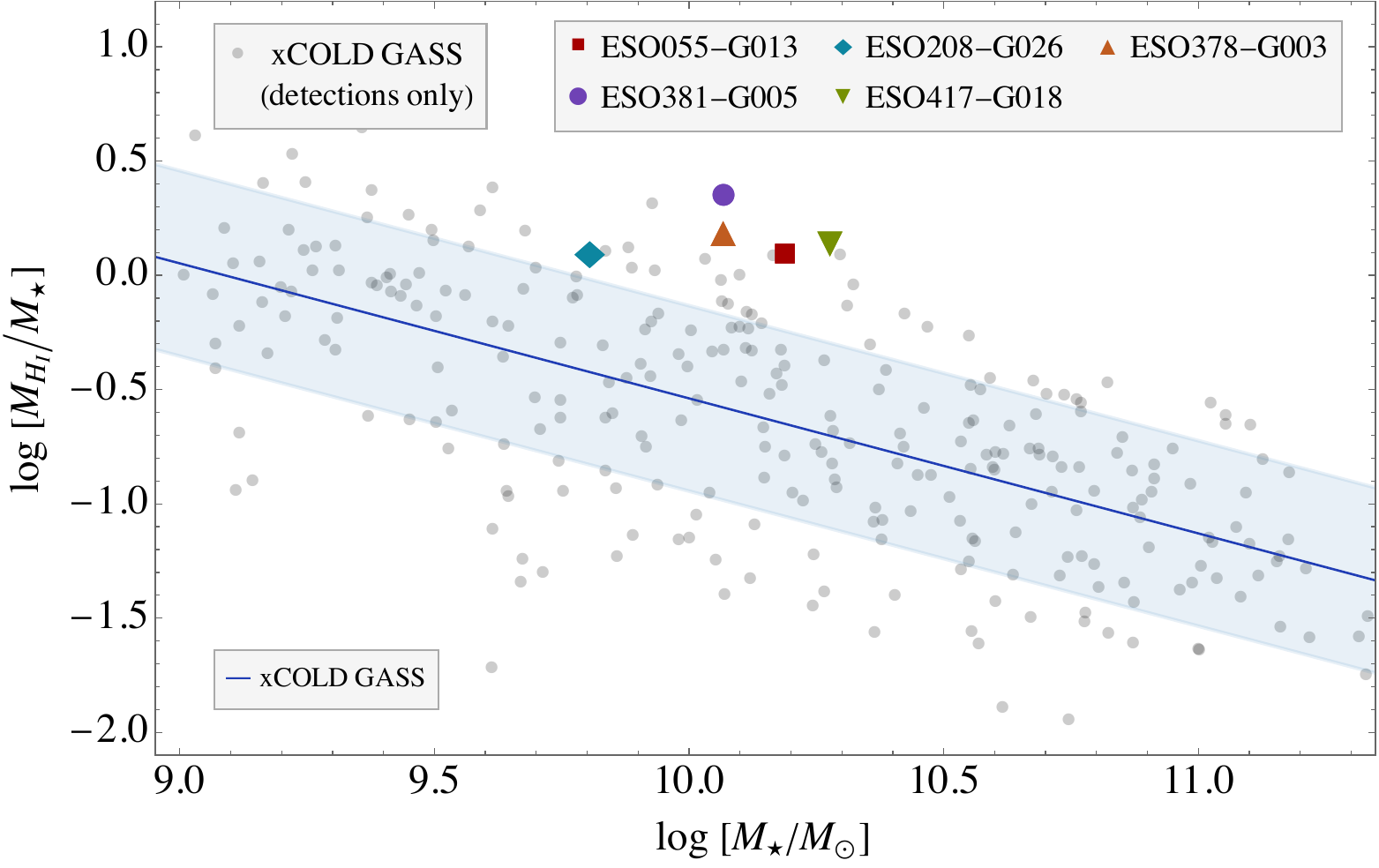}
  \includegraphics[width = 6.5cm]{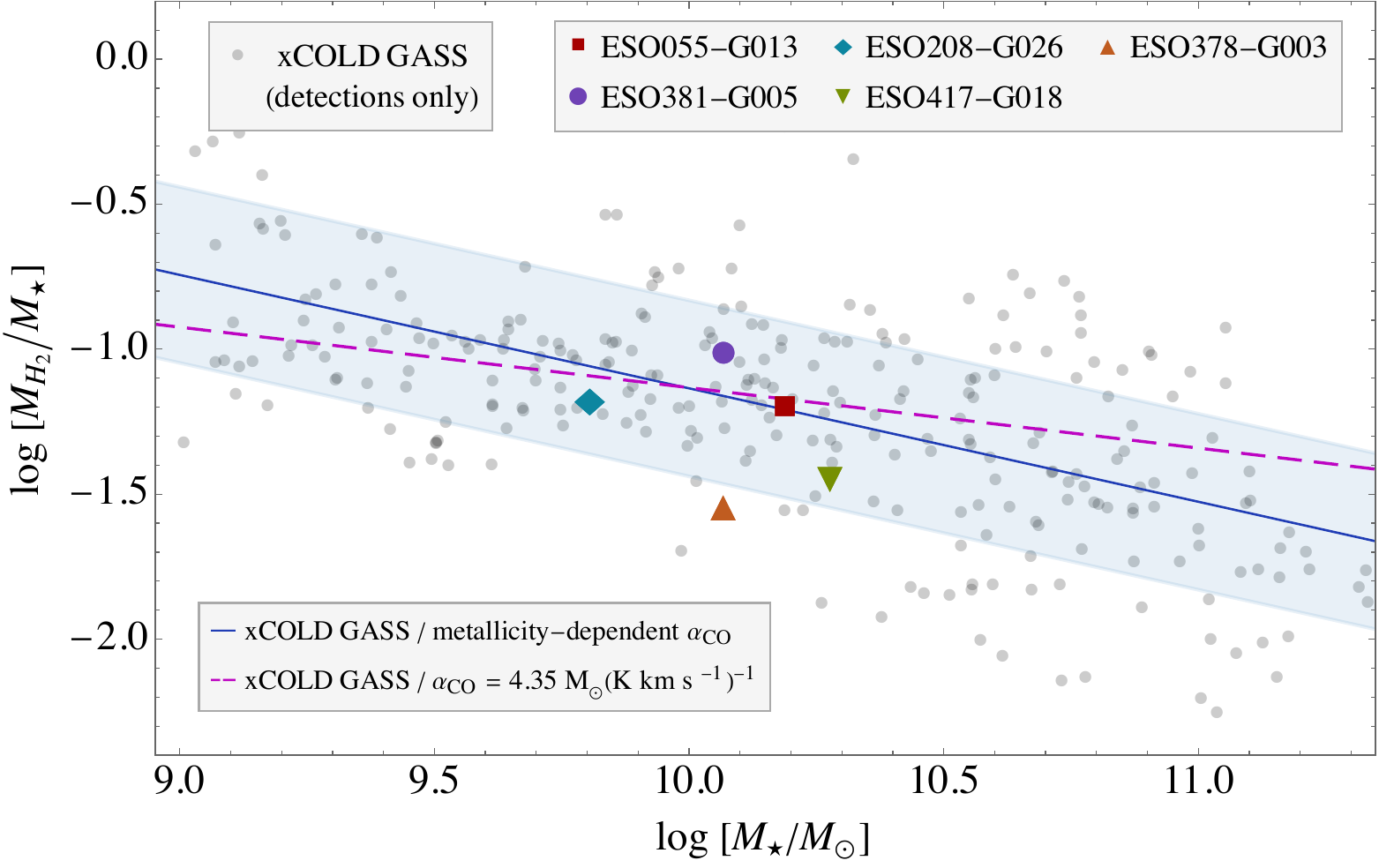}
  \includegraphics[width = 6.5cm]{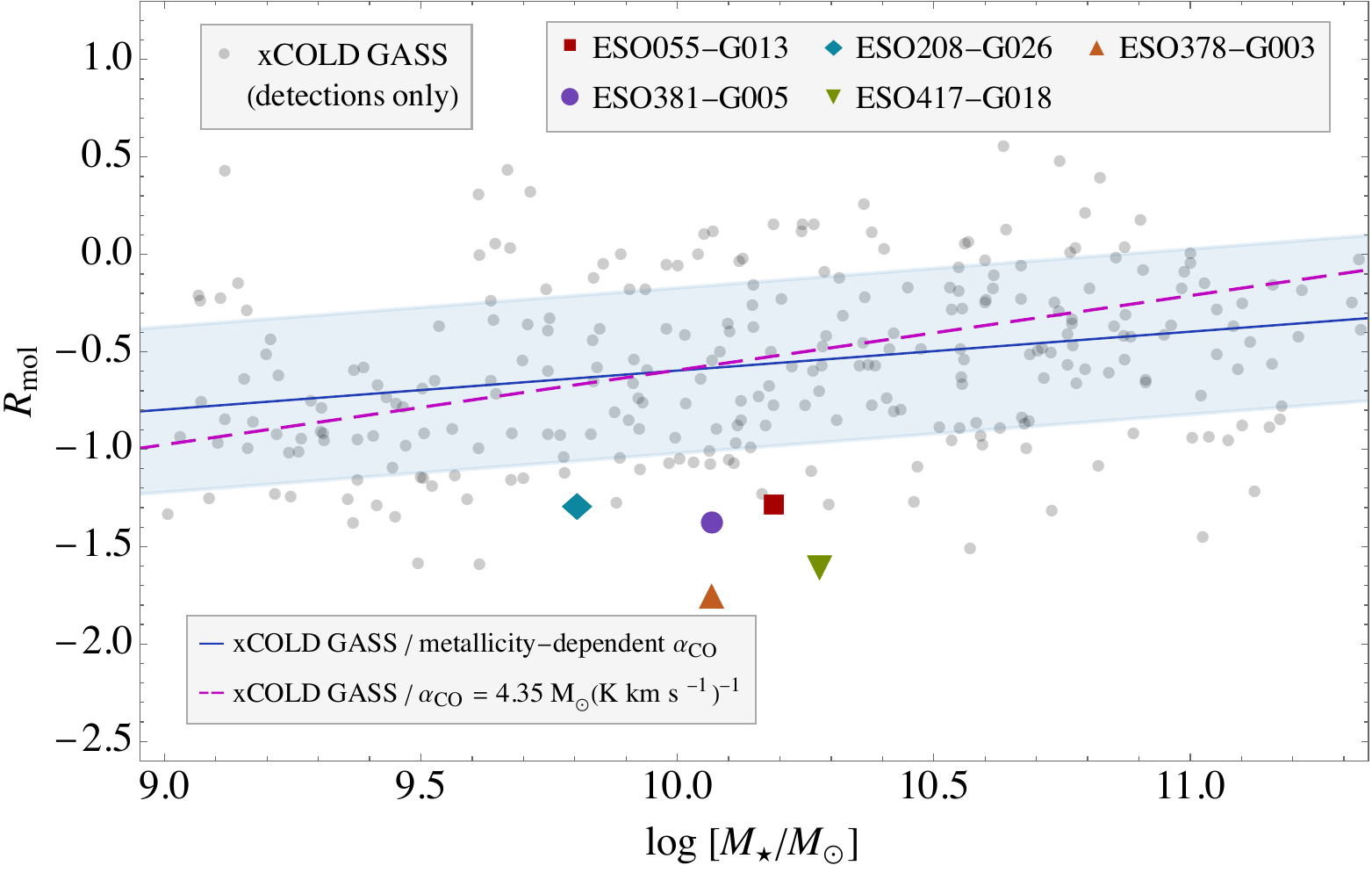}
  \caption{Scaling relations of \HI\ to M$_{\star}$ ratio (upper left), \Htwo\ to M$_{\star}$ ratio (upper right) and \HI\ to \Htwo\ ratio (lower left) as a function of stellar mass. The coloured markers indicate our sample galaxies, while the grey circles are measurements from the xCOLD GASS sample. The solid lines indicate the fitted \HI\ fraction relation and the metallicity dependent $H_2$ fraction relations, while the dashed lines indicate the metallicity independent $H_2$ fraction relations. The shaded regions indicate $\pm \sigma$ from the fitted relations.}
  \label{fig:scaling_relations}
\end{figure}

Figure~\ref{fig:scaling_relations} shows the \HI-$M_{\star}$ and \Htwo-$M_{\star}$ scaling relations as a function of stellar mass ($M_{\star}$). Data from the xCOLD GASS survey is marked with grey circles and data from our sample is marked with coloured symbols. We also included fitted lines and the $1\sigma$ shaded region obtained from \cite{Saintonge2017}. The scaling relations clearly show the relative excess of \HI\ in our sample galaxies. Except ESO378-G003, which has a relatively low \Htwo\ mass, all galaxies have a regular \Htwo\ content. In terms of \Htwo/\HI\ content all galaxies have a low molecular gas fraction. Comparing the \Htwo\ mass with the star formation rate we get \Htwo\ depletion times for the sample between 0.4 - 1.4 $\times 10^9$ yr, which are in the same range as the galaxies observed by \cite{Hallenbeck2016}, supporting inefficient \HI\ to \Htwo\ conversion.

\section{Conclusions}

Based on the presented gas scaling relations, the \hix\ galaxies seem to have a regular \Htwo\ mass compared to their stellar mass, and a low molecular gas fraction. Comparing the \Htwo\ mass with the star formation rate of the sample we get \Htwo\ depletion times that are lower compared to those of nearby galaxies. The \co\ moment maps indicate a regularly rotating \co\ disk covering the inner part of the stellar disk and an elevated velocity dispersion. The relatively high velocity dispersion may indicate higher turbulence compared to nearby galaxies. Taking into account that the \HI\ distribution of all five sample galaxies is disturbed in the outer part (see Figure~\ref{fig:Maps}), we interpret the gas richness of the \hix\ sample as a result of recent interactions with neighbouring galaxies that may have resulted in the acquisition of \HI\ or in increasing the velocity dispersion of the gas. 

We note that these are our preliminary conclusions and we are going to expand the data analysis of these galaxies in an upcoming publication by Capa et al. (in prep.).

\end{document}